\documentclass[aps,preprint,showpacs,preprintnumbers,amsmath,amssymb]{revtex4}
\usepackage{amsmath,mathrsfs,amsbsy,color,graphicx,bm,amsthm,amsfonts}
\usepackage{units}
\usepackage{bbm}
\usepackage{times}
\usepackage{dcolumn}
\usepackage{mathrsfs}
\usepackage{amsmath,amssymb,epsfig}

\begin{document}

\title{Quantum entanglement for continuous variables sharing in an
expanding spacetime }
\author{Wen-Mei Li$^1$, Rui-Di Wang$^1$, Hao-Yu Wu$^1$, Xiao-Li Huang$^1$\footnote{ huangxiaoli1982@foxmail.com}, Hao-Sheng Zeng$^2$\footnote{hszeng@hunnu.edu.cn },Shu-Min Wu$^1$\footnote{smwu@lnnu.edu.cn (corresponding author)}}
\affiliation{$^1$ Department of Physics, Liaoning Normal University, Dalian 116029, China\\
$^2$ Department of Physics, Hunan Normal University, Changsha 410081, China
}


\begin{abstract}
Detecting the structure of spacetime with quantum technologies has always been one of the frontier topics of relativistic quantum information. Here, we analytically study the generation and redistribution of Gaussian entanglement of the scalar fields in an expanding spacetime. We consider a two-mode squeezed state via a Gaussian amplification channel that corresponds to the time-evolution of the state from
the asymptotic past to the asymptotic future. Therefore, the dynamical entanglement of the Gaussian state in an expanding universe encodes historical information about the underlying spacetime structure, suggesting a promising application in observational cosmology. We find that quantum entanglement is more sensitive to the expansion rate than the expansion volume.
According to the analysis of quantum entanglement, choosing the particles with the smaller momentum and the optimal mass is a better way to extract information about the expanding universe. These results can guide  the simulation of the expanding universe in quantum systems.
\end{abstract}

\vspace*{0.5cm}
\maketitle
\section{Introduction}
Quantum entanglement, predicted by Schr\"{o}dinger in 1935,
plays an important role in quantum information
theory.  It is considered to be a resource for quantum information tasks, such as computational task, quantum teleportation, quantum communication,  quantum control, and quantum simulations \cite{L1,L2,L3,L4}.
Recently, the influence of gravitational effects on quantum entanglement
 has received considerable attention.
Therefore, many efforts have been expended on the study of
quantum entanglement of the field modes in a noninertial frame, in the black hole, and in de Sitter space \cite{L5,L6,LA1,LA2,L7,L8,L9,L10,L11,L12,L13,L14,L15,L16,L17,L18,L19,L20,OL1,OL2,OL3,OL4,OL5,OL6,OL7,OL8,OL9,OL10,LDF1,LDF2,LDF3,LDF4,LDF5}.
Besides, many papers have studied the influences of different spacetime on quantum entanglement and explored the properties of spacetime using  quantum entanglement.
It is clear that these investigations not only contribute to understanding the key questions  about  quantum information and exploring the structure of spacetime, but also play a central role in the study of the information paradox of the black hole and the entanglement entropy \cite{L21,L22,L23,L24}.

In this work, we study the redistribution of bipartite entanglement for continuous variables of the scalar fields in the background of an expanding universe.
We initially consider a two-mode
squeezed Gaussian state shared by Alice and Bob in the asymptotic past.
A vacuum state in the asymptotic past evolves into a
thermal state through the expansion of the universe in the asymptotic
future \cite{L25,L26,L27,L28}. In a quantum information scenario, such a process of the expansion of the universe can be described as a Gaussian  channel acting on a two-mode
squeezed Gaussian state.
There are two reasons why I choose the two-mode
squeezed Gaussian state: firstly, it is a typical continuous variable entangled state, which
approximates to an arbitrarily good extent the EPR pair; secondly,
the state can be produced in the
laboratory and exploited for any current realization of quantum information with continuous variables  \cite{L29,L30}.
In the standard cosmology model, the dynamical quantum entanglement  for continuous variables acts as witnesses in the history of the
universe from the era of big-bang nucleosynthesis to the era of large-scale structure formation and can encode historical information about the underlying spacetime structure, which suggests a promising application in observational cosmology. Thus, studying the behavior of quantum entanglement is crucial to understanding the history of our universe and its fate.

A two-mode squeezed Gaussian state in the asymptotic past becomes a four-mode Gaussian state in the asymptotic future. Therefore, we
evaluate not only the initial bipartite entanglement as influenced by
the expansion of the universe, but also, remarkably, the multipartite  entanglement,
which arises from two bosonic modes and two antibosonic modes.
The result shows that the bipartite entanglement initially prepared in the asymptotic past is
exactly redistributed into  four-partite correlations in the asymptotic future, as a consequence of the monogamy constraints on quantum entanglement distribution.
On the other hand, we understand the properties of the expanding spacetime through quantum entanglement for continuous variables. Our results may guide  the  cosmological observations and  the  simulation of the expanding
spacetime in quantum systems \cite{oL45m,oL45m1,oL45m1po}.

The paper is organized as follows. In Sec. II, we briefly introduce the measures of bipartite entanglement for continuous variables and the Coffman-Kundu-Wootters inequality. In Sec. III,
we discuss how the expansion of the universe is
described by a Gaussian channel. In Sec. IV, we study the redistribution of the two-mode Gaussian  entanglement and the generated $1\rightarrow3$ entanglement under the influence of the expansion of the underlying spacetime. The last section is devoted to a brief conclusion.

\section{Quantifying entanglement for continuous variables by the R\'{e}nyi-2 entropy \label{GSCDGE}}
The set of Gaussian states by the definition is the set of states with quasi-probability distributions and Gaussian characteristic functions in the quantum phase space.
The properties of the Gaussian state are entirely determined by  the first and second
statistical moments of the quadrature operators.
As the first moments can be arbitrarily
adjusted by the local unitary operations, which keep all informationally relevant properties
(such as entropy or any measure of correlations) invariant, we can adjust them to
be zero. Therefore, the second moments for describing
Gaussian state become the unique elements. Based on our research,
we initially consider a two-mode Gaussian state $\rho_{AB}$
shared by Alice and Bob.
We define a vector of quadrature operators as $\hat R = ( \hat{X}_1, \hat{P}_1,...,  \hat{X}_n, \hat{P}_n)^{\sf T}$, which satisfies the canonical commutation
relations $[{{{\hat R}_k},{{\hat R}_l}} ] = i{\Omega _{kl}}$, with the symplectic matrix $\Omega = {{\ 0\ \ 1}\choose{-1\ 0}}^{\oplus{n}}$ \cite{L31,L32,L33}.
The elements of a covariance matrix (CM) $\sigma_{AB}$ can be defined as ${\sigma _{ij}} = {\rm Tr}\big[ {{{\{ {{{\hat R}_i},{{\hat R}_j}} \}}_ + }\ {\rho _{AB}}} \big]$.
The CM $\sigma_{AB}$ of the Gaussian state can be put into a block form
\begin{equation}\label{w1}
 \sigma_{AB}= \left(
                      \begin{array}{cc}
                       \mathcal{A} & \mathcal{C} \\
                        \mathcal{C}^{\rm T} & \mathcal{B} \\
                      \end{array}
                    \right).
\end{equation}
For a physical Gaussian state, its CM $\sigma_{AB}$ must satisfy the uncertainty relation
\begin{eqnarray}\label{w2}
{\sigma _{AB}} + i\Omega \ge 0.
\end{eqnarray}

R\'{e}nyi-$\alpha$ entropies consist of a powerful family of additive entropies,
which can  provide a measure of quantum information. R\'{e}nyi-$\alpha$ entropies
are defined as \cite{L34,L35}
\begin{equation}\label{w3}
S_\alpha(\rho)=\frac{1}{1-\alpha}\ln {\rm tr}(\rho^\alpha).
\end{equation}
In the limit $\alpha\rightarrow1$, it reduces to the von Neumann entropy.
For the case $\alpha=2$, the special R\'{e}nyi-2 entropy can be calculated very easily
\begin{equation}\label{w4}
S_2(\rho)=-\ln {\rm tr}(\rho^2)=\frac{1}{2}\ln(\det \sigma),
\end{equation}
where $\sigma$ is the CM of the Gaussian state with density matrix $\rho$.
For arbitrary Gaussian states, R\'{e}nyi-$\alpha$ entropies fulfill the strong subadditivity
inequality.  This allows us to define relevant Gaussian measures of information and
entanglement quantities under a unified approach.

For a $N$-mode bipartite Gaussian state with CM $\sigma_{AB}$ in Eq.(\ref{w1}), the R\'{e}nyi-2 quantum entanglement $\mathcal{E}({\sigma_{AB}})$,  quantifying the quantum entanglement between Alice and Bob, can be defined as \cite{L36}
\begin{equation}\label{w5}
    \mathcal{E}({\sigma_{AB}})=\inf_{\gamma_{AB}}\frac{1}{2}\ln(\det\gamma_{A}).
\end{equation}
For a pure Gaussian state, the minimum
is saturated by $\sigma_{AB}=\gamma_{AB}$, so that $\mathcal{E}({\sigma_{AB}})=\frac{1}{2}\ln(\det \gamma_{A})$,
where $\gamma_{A}$ is the reduced CM for Alice. For a mixed state,
the minimization is over pure N-mode Gaussian states with CM
$\gamma_{AB}$ which satisfies $0<\gamma_{AB}\leq\sigma_{AB}$ and $\det\gamma_{AB}=1$.
[For two real symmetric matrices $\mathbf{M}$ and $\mathbf{N}$, $\mathbf{M}\geq \mathbf{N}$ means that the matrix $\mathbf{M}-\mathbf{N}$ has all non-negative eigenvalues.]
For a special class of two-mode Gaussian states, the closed formulae $\mathcal{E}$
can be obtained by the same procedure of entanglement of formation \cite{L37}.

In multipartite systems, quantifying entanglement is generally
very involved. Unlike classical correlations, quantum entanglement is monogamous, which means that it cannot be freely shared among multiple subsystems
of a composite system. So far, this fundamental
constraint on quantum entanglement sharing has been
mathematically demonstrated not only for arbitrary systems of
qubits within the discrete-variable scenario but also for a
special case of two qubits and an infinite-dimensional system and for all N-mode Gaussian states within the continuous variables  scenario.
In the general case of a quantum state distributed among $N$ parties,
the monogamy constraint can be presented in the form of the Coffman-
Kundu-Wootters inequality \cite{L36}
\begin{equation}\label{w6}
    \mathcal{E}_{S_i|(S_1...S_{i-1}S_{i+1}...S_N)}\geq \sum_{j\neq i}^N\mathcal{E}_{S_i|S_j},
\end{equation}
where the multipartite system has $N$ subsystems $S_k(k=1,...,N)$, each owned by a corresponding party, and $\mathcal{E}$ is a proper quantification of bipartite entanglement.
The left-hand side of inequality (\ref{w6}), $\mathcal{E}_{S_i|(S_1...S_{i-1}S_{i+1}...S_N)}$,
can quantify the bipartite entanglement between a probe subsystem $S_i$ and the remaining $N-1$ subsystems. The right-hand side of inequality (\ref{w6}), $\sum_{j\neq i}^N\mathcal{E}_{S_i|S_j}$, can  quantify the total
bipartite entanglements between subsystem $S_i$ and each one of the other
subsystems $S_{j\neq i}$ in the reduced states. The nonnegative difference between these two quantum entanglements, which is minimized
over all choices of the probe subsystem, is known
as the residual multipartite entanglement.
It quantifies the quantum entanglements not encoded in pairwise
form, so it includes all manifestations of genuine N-partite
entanglement.

\section{The expansion of the universe described by Gaussian channel \label{GSCDGE}}
Let us start with a 1+1 dimensional Robertson-Walker expanding universe with the metric $ds^2=dt^2-[a(t)]^2 dx^2$,
where $a(t)$ is the scale factor.
With the conformal time $\eta$ relating to the
cosmological time $t$ by $\eta=\int_0^t \frac{\text{d}\tau}{a(\tau)}$,
the metric of the Robertson-Walker expanding universe can be rewritten as \cite{L25,L26,L27,L28}
\begin{equation}\label{w7}
ds^2=[a(\eta)]^2(d\eta^2- dx^2)\,.
\end{equation}
Here, the conformal scale factor takes the form
\begin{equation}\label{w8}
[a(\eta)]^2=1+\epsilon(1+\tanh(\sigma\eta)),
\end{equation}
where the parameters $\epsilon$ and $\sigma$ characterize the volume and
the rapidity of the expansion, respectively.
It is obvious that the spacetime is flat in the distant past and
the far future corresponding to the metric $ds^2=d\eta^2- dx^2$ when $\eta\rightarrow -\infty$
and  the metric $ds^2=(1+2\epsilon)(d\eta^2- dx^2)$ when $\eta\rightarrow +\infty$, respectively.
Therefore, the timelike Killing vector and the particle content of the
field are defined in these two limits.

A real scalar field $\Phi(x,\eta)$ in the Robertson-Walker expanding spacetime obeys the Klein-Gordon equation
\begin{equation}\label{w9}
(\Box+ m^2)\Phi  =0,
\end{equation}
where $\Box=\frac{1}{\sqrt{|g|}}\partial_\mu \sqrt{|g|} g^{\mu\nu}\partial_\nu$.
Having  solved the Klein-Gordon equation at the limits of $\eta\rightarrow \pm\infty$,
we obtain a set of modes $u^{\text{in}}$ in the distant past (``in"  region)
and a set of modes $u^{\text{out}}$ in the far future (``out" region).
Using the inner product, the Bogoliubov transformations between the modes $u^{\text{in}}_k$ and  $u^{\text{out}}_k$ take the form
\begin{equation}\label{w10}
u^{\text{in}}_k(x,\eta)=\alpha_k  u^{\text{out}}_k(x,\eta)+\beta_k u^{{\text{out}*}}_{-k}(x,\eta)\,,
\end{equation}
where the Bogoliubov coefficients are given by
\begin{eqnarray}\label{w11}
\alpha_k&=\sqrt{\frac{\omega_{\text{out}}}{\omega_{\text{in}}}}\frac{\Gamma([1-(i\omega_{\text{in}}/\sigma)])\Gamma(-i\omega_{\text{out}}/\sigma)}{\Gamma([1-(i\omega_{+}/\sigma)])\Gamma(-i\omega_{+}/\sigma)}\,, \\[3mm]
\label{betak}\beta_k&=\sqrt{\frac{\omega_{\text{out}}}{\omega_{\text{in}}}}\frac{\Gamma([1-(i\omega_{\text{in}}/\sigma)])\Gamma(i\omega_{\text{out}}/\sigma)}{\Gamma([1+(i\omega_{-}/\sigma)])\Gamma(i\omega_{-}/\sigma)}\,,
\end{eqnarray}
with  $\Gamma$ being the gamma function, $
\omega_{\text{in}}=\sqrt{k^2+m^2}\,, \qquad
\omega_{\text{out}}=\sqrt{k^2+m^2(1+2\epsilon)}\,, \qquad
\omega_{\pm}=\frac12(\omega_{\text{out}}\pm \omega_{\text{in}})\,.
$
Through simple calculation, the Bogoliubov coefficients satisfy $|\alpha_k|^2-|\beta_k|^2=1$.
For convenience, we define $ \theta_k^2=|\frac{\beta_k}{\alpha_k}|^2=
\frac{\sinh^2(\pi\frac{\omega_-}{\sigma})}
{\sinh^2(\pi\frac{\omega_+}{\sigma})}$ and easily obtain
\begin{eqnarray}\label{w13}
|\alpha_k|^2=\frac{1}{1-\theta_k^2}, |\beta_k|^2=\frac{\theta_k^2}{1-\theta_k^2},
\end{eqnarray}
where $|\beta_k|^2$ equals the average number of particles
created at  ``out" mode $k$. Hence, $\theta_k^2\rightarrow 0$  means that
the average number of particles of the mode $k$ is vanishing, and
$\theta_k^2\rightarrow 1$ means that
the average number of particles of the mode $k$ approaches infinity.

The annihilation and creation operators satisfy
\begin{equation}\label{w14}
b_{\text{in},k}=\alpha_k^*b_{\text{out},k}-\beta_k^*b_{\text{out},{-k}}^\dagger,
\end{equation}
\begin{equation}\label{w15}
b_{\text{in},k}^\dagger=\alpha_kb_{\text{out},k}^\dagger-\beta_kb_{\text{out},{-k}} ,
\end{equation}
where $b_{\text{in},k}$ and $b_{\text{in},k}^\dagger$ are the bosonic annihilation
and creation operators acting on the states
in the asymptotic past, $b_{\text{out},k}$ and $b_{\text{out},k}^\dagger$ are
the the bosonic annihilation and creation operators acting
on the states in the asymptotic future, and $b_{\text{out},-k}$ and $b_{\text{out},-k}^\dagger$ are the the antibosonic annihilation and creation operators, respectively. We use $b_{\text{in},k}|0_k\rangle_{\text{in}}=0$ to find the relation between the ``in" vacuum state and the ``out" vacuum state.
If substituting $b_{\text{in},k}$ with  Eq.(\ref{w14}), we obtain
\begin{equation}\label{w16}
(\alpha_k^*b_{\text{out},k}-\beta_k^*b_{\text{out},{-k}}^\dagger)|0_k\rangle_{\text{in}}=0.
\end{equation}
According to the the normalization condition, the ``in" vacuum state  can be expressed in the asymptotic future as
\begin{equation}\label{w17}
|0_k\rangle_{\text{in}}=\sum_{n=0}^\infty
A_n
|n_k\rangle_{\text{out}} |n_{-k}\rangle_{\text{out}},
\end{equation}
where $A_n=\sqrt{1-\theta_k^2}(\frac{\beta^*_k}{\alpha^*_k})^n$, $n_k$ denotes the boson number, and $n_{-k}$ denotes the antiboson number.
This means that an initial vacuum state $|0_k\rangle_{\text{in}}$
evolves into a two-mode squeezed state in the asymptotic future.
By rotating the squeezing angle and giving up  the phase angle, we obtain \cite{L38,L39,L40,L41}
\begin{equation}\label{w18}
|0_k\rangle_{\text{in}}=\sqrt{1-\theta_k^2}\sum_{n=0}^\infty
\theta_k^n
|n_k\rangle_{\text{out}} |n_{-k}\rangle_{\text{out}}=U_k|0_k\rangle|0_{-k}\rangle,
\end{equation}
where $U_k=\exp[r_k(b_{\text{out},{k}}^\dagger b_{\text{out},{-k}}^\dagger-b_{\text{out},{k}} b_{\text{out},{-k}})]$ is a two-mode squeezing
operator. The squeezing parameter $r_k$ is defined as $\cosh(r_k)=|\alpha_k|$.
It is worth emphasizing that the squeezing operator $U_k$ is a Gaussian operation, which
preserves the Gaussianity from the input states. Therefore, Eq.(\ref{w18}) shows that  the expansion of a Robertson-Walker spacetime can be described by a Gaussian (a bosonic amplification) channel. In the phase space,
the two-mode squeezing operator $U_k$ corresponds to the symplectic transformation
\begin{eqnarray}\label{w19}
 S_k=\frac{1}{\sqrt{1-\theta_k^2}} \left(
                      \begin{array}{cc}
                        I_{2} & \theta_k Z_2 \\
                       \theta_k Z_2 & I_{2} \\
                      \end{array}
                    \right),
\end{eqnarray}
where $I_{2}$ denotes the unity matrix in $2\times 2$ space, and $Z_2$ denotes the third Pauli matrix.

\section{The effect of the expanding universe on Gaussian entanglement \label{GSCDGE}}
In this paper, we initially consider a pure two-mode squeezed Gaussian state shared by Alice and Bob  in the distant past. It has covariance matrix \cite{L40}
\begin{eqnarray}\label{w20}
 \sigma_{AB}^{\rm in}= \left(\!\!\begin{array}{cc}
\cosh(2s)I_{2} & \sinh(2s)Z_{2}\\
\sinh(2s)Z_{2} & \cosh(2s)I_{2}
\end{array}\!\!\right),
\end{eqnarray}
where $s$ is the squeezing parameter.
If Alice and Bob undergo the expanding universe associating to the symplectic transformation in Eq.(\ref{w19}), the initial two-mode squeezed state $\sigma_{AB}^{\rm in}$ becomes a four-mode Gaussian state in the asymptotic future.
In other words, the initial two-mode squeezed state  in the asymptotic past is transformed, via the expansion of a Robertson-Walker spacetime, to a four-mode Gaussian state in the asymptotic future. Therefore, a complete description
of the quantum system involves four modes: the bosonic mode $A$  described by Alice; the bosonic mode $B$  described by Bob; the antibosonic mode $\bar A$
described by anti-Alice; the antibosonic mode $\bar B$  described by anti-Bob. The covariance matrix describing the complete system thus becomes \cite{L40}
\begin{eqnarray}\label{w21}
\sigma_{AB \bar A \bar B}^{\rm out}=\big[S_{A,\bar A} \oplus  S_{B,\bar B}\big] \big[\sigma_{AB}^{\rm in} \oplus I_{\bar A\bar B}\big] \big[S_{A,\bar A} \oplus  S_{B,\bar B}\big]\,^{\sf T},
\end{eqnarray}
where $S_{A,\bar A}$ and $S_{B,\bar B}$ given by Eq.(\ref{w19}) are the phase space representation of the two-mode squeezing operation, and $I_{\bar A\bar B}$ denotes the $4\times 4$ identity matrix.
\subsection{Bipartite Gaussian entanglement}
Because Alice and Bob cannot detect the antibosonic modes, we should take the
trace over the modes $\bar A$ and $\bar B$. By performing this operation
on  Eq.(\ref{w21}), we obtain the reduced state between the modes $A$ and $B$
\begin{eqnarray}\label{w22}
 \sigma_{AB}^{\rm out}=\frac{1}{1-\theta_k^2} \left(
                      \begin{array}{cc}
                        [\cosh(2s)+\theta_k^2]I_{2} & \sinh(2s) Z_2 \\
                       \sinh(2s) Z_2 &  [\cosh(2s)+\theta_k^2]I_{2} \\
                      \end{array}
                    \right).
\end{eqnarray}
Employing Eq.(\ref{w5}), we can obtain an analytic
expression of Gaussian entanglement
\begin{equation}\label{w23}
    \mathcal{E}(\sigma_{AB}^{\rm out})=\ln \bigg\{-\frac{\cosh(2s)(\theta_k^4-\theta_k^2+1)+\theta_k^2[\sinh(2s)(\theta_k^2-1)+1]}
    {(\theta_k^2-1)(\cosh(s)+\sinh(s))[\cosh(s)(\theta_k^2+1)+\sinh(s)(\theta_k^2-1)]}\bigg\}.
\end{equation}
From Eq.(\ref{w23}), we can see that quantum entanglement $\mathcal{E}(\sigma_{AB}^{\rm out})$ depends not only on the squeezing parameter $s$, but also on the volume $\epsilon$ and
the rapidity $\sigma$ of the expanding universe, meaning that quantum entanglement  encodes historical information about the cosmological parameters.

\begin{figure}
\begin{minipage}[t]{0.5\linewidth}
\centering
\includegraphics[width=3.0in,height=5.2cm]{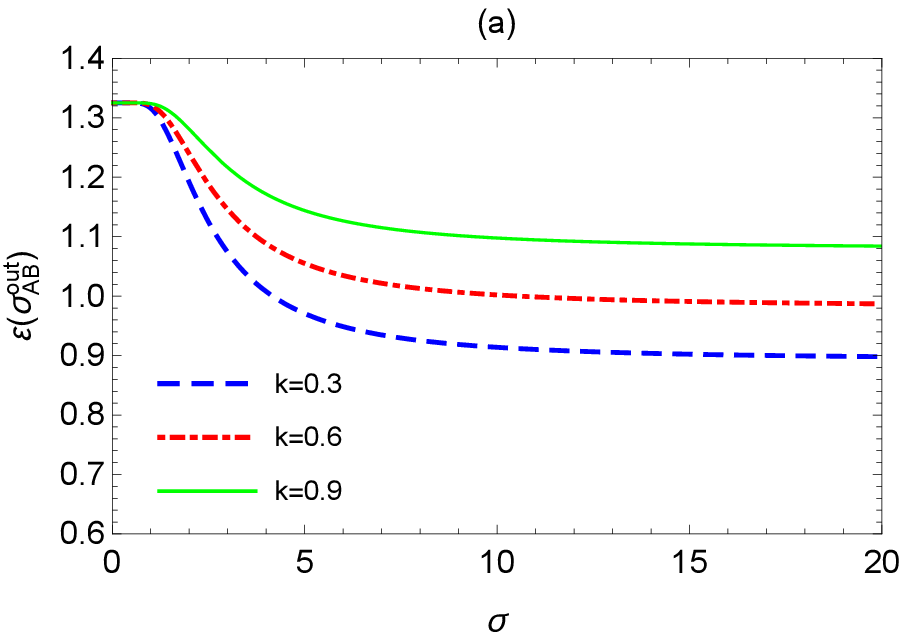}
\label{fig1a}
\end{minipage}%
\begin{minipage}[t]{0.5\linewidth}
\centering
\includegraphics[width=3.0in,height=5.2cm]{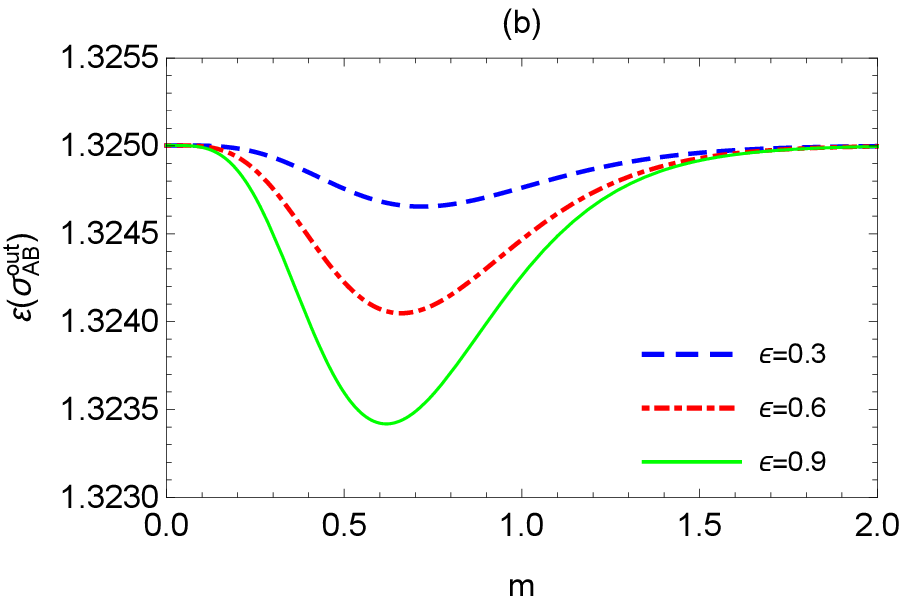}
\label{fig1c}
\end{minipage}%
\caption{(a) Quantum entanglement $\mathcal{E}(\sigma_{AB}^{\rm out})$ between the modes $A$ and $B$ as a function of the expansion rate $\sigma$ for different momentums $k$, with the fixed values of $m=\epsilon=s=1$. (b) Quantum entanglement $\mathcal{E}(\sigma_{AB}^{\rm out})$ as a function of the mass $m$ for different expansion volumes $\epsilon$, with the fixed values of $k=\sigma=s=1$.}
\label{Fig1}
\end{figure}

In Fig.\ref{Fig1} (a), we plot the Gaussian entanglement $\mathcal{E}(\sigma_{AB}^{\rm out})$ between Alice and Bob as a function of the expansion rate $\sigma$ for different momentums $k$. Fig.\ref{Fig1} (b) shows how the mass $m$
influences the Gaussian entanglement $\mathcal{E}(\sigma_{AB}^{\rm out})$ for different expansion volumes $\epsilon$. From Fig.\ref{Fig1} (a), we can see that quantum entanglement $\mathcal{E}(\sigma_{AB}^{\rm out})$ is a monotonically decreasing function of the expansion rate $\sigma$ and a monotonically increasing function of the momentum $k$.
We find that as the expansion rate $\sigma$ increases, quantum entanglement decreases to an asymptotic value dependent on the momentum $k$.
At the limit of $k\rightarrow\infty$, we obtain
$$\lim_{k\rightarrow\infty}\theta_k=0, \qquad \lim_{k\rightarrow\infty}\mathcal{E}(\sigma_{AB}^{\rm out})=\ln[\cosh(2s)],$$
where $\ln[\cosh(2s)]$ is the initial entanglement in Eq.(\ref{w20}).
Therefore, we cannot extract any information about the spacetime at the limit of $k\rightarrow\infty$. Fig.1 (b) shows that quantum entanglement
$\mathcal{E}(\sigma_{AB}^{\rm out})$  decreases with the increase of the expansion volume $\epsilon$.
We can also see that quantum entanglement between the modes $A$ and $B$ first decreases to the minimum value and then increases to the initial value with the growth of
the mass $m$. Therefore, quantum entanglement of the bosonic fields with the larger mass $m$ is insensitive to the expansion volume $\epsilon$.
This means that choosing the bosons  with the appropriate mass makes quantum entanglement more sensitive to the cosmological parameters.

We  calculate quantum entanglements in all possible
bipartite divisions of the four-mode quantum system to explore the distribution of Gaussian entanglement in an expanding spacetime. Firstly,
tracing over the modes in $A$ and $B$, we obtain the covariance matrix $\sigma_{\bar A \bar B}^{\rm out}$ between the modes $\bar A$ and $\bar B$
\begin{eqnarray}\label{w24}
 \sigma_{\bar A \bar B}^{\rm out}=\frac{1}{1-\theta_k^2} \left(
                      \begin{array}{cc}
                        [\theta_k^2\cosh(2s)+1]I_{2} & \theta_k^2\sinh(2s) Z_2 \\
                       \theta_k^2 \sinh(2s) Z_2 &  [\theta_k^2\cosh(2s)+1]I_{2} \\
                      \end{array}
                    \right).
\end{eqnarray}
Secondly, taking the trace over the modes $B$ and $\bar A$,
we get the covariance matrix between  Alice and anti-Bob
\begin{eqnarray}\label{w25}
 \sigma_{ A \bar B}^{\rm out}=\frac{1}{1-\theta_k^2} \left(
                      \begin{array}{cc}
                        [\cosh(2s)+\theta_k^2]I_{2} & \theta_k\sinh(2s) I_2 \\
                       \theta_k \sinh(2s) I_2 &  [\theta_k^2\cosh(2s)+1]I_{2} \\
                      \end{array}
                    \right).
\end{eqnarray}
$\sigma_{ \bar AB }^{\rm out}=\sigma_{ A \bar B}^{\rm out}$ can be directly obtained by a simple calculation. According to Eq.(\ref{w5}), we find $\mathcal{E}(\sigma_{\bar A \bar B}^{\rm out})=\mathcal{E}(\sigma_{A \bar B}^{\rm out})=0$,
meaning that  the expansion of the universe cannot generate quantum entanglement between  anti-Alice and anti-Bob (or  Alice and anti-Bob).

Finally, we fix our eyes on quantum entanglement between the modes $A$  and $\bar A$. Tracing over the modes $B$ and $\bar B$, we obtain the covariance matrix $\sigma_{ A \bar A}^{\rm out}$ for Alice and anti-Alice
\begin{eqnarray}\label{w26}
 \sigma_{ A \bar A}^{\rm out}=\frac{1}{1-\theta_k^2} \left(
                      \begin{array}{cc}
                        [\cosh(2s)+\theta_k^2]I_{2} & 2\theta_k\cosh^2(s) Z_2 \\
                       2\theta_k \cosh^2(s) Z_2 &  [\theta_k^2\cosh(2s)+1]I_{2} \\
                      \end{array}
                    \right).
\end{eqnarray}
Using Eqs.(\ref{w5}) and (\ref{w26}), we obtain an
analytic expression of quantum entanglement between the modes $A$ and $\bar A$
\begin{equation}\label{w27}
    \mathcal{E}(\sigma_{A\bar A}^{\rm out})=\ln \bigg\{\frac{1+\theta_k^2}{1-\theta_k^2}  \bigg\}.
\end{equation}
Unlike the Gaussian entanglement $\mathcal{E}(\sigma_{AB}^{\rm out})$, the Gaussian entanglement $\mathcal{E}(\sigma_{A\bar A}^{\rm out})$ does not depend on the initial squeezing parameter $s$. What we need to notice is that $\sigma_{ B \bar B}^{\rm out}$ is equal to $\sigma_{ A \bar A}^{\rm out}$.

\begin{figure}
\begin{minipage}[t]{0.5\linewidth}
\centering
\includegraphics[width=3.0in,height=5.2cm]{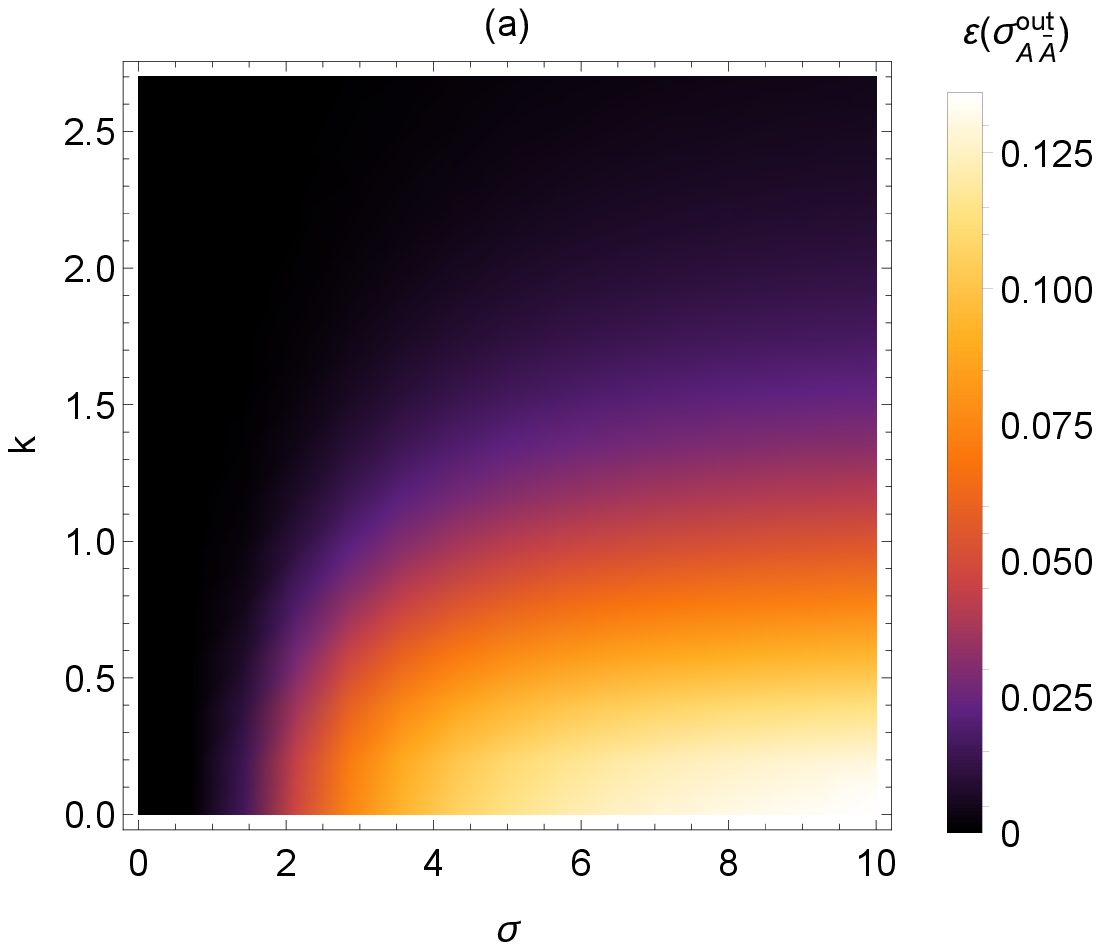}
\label{fig1a}
\end{minipage}%
\begin{minipage}[t]{0.5\linewidth}
\centering
\includegraphics[width=3.0in,height=5.2cm]{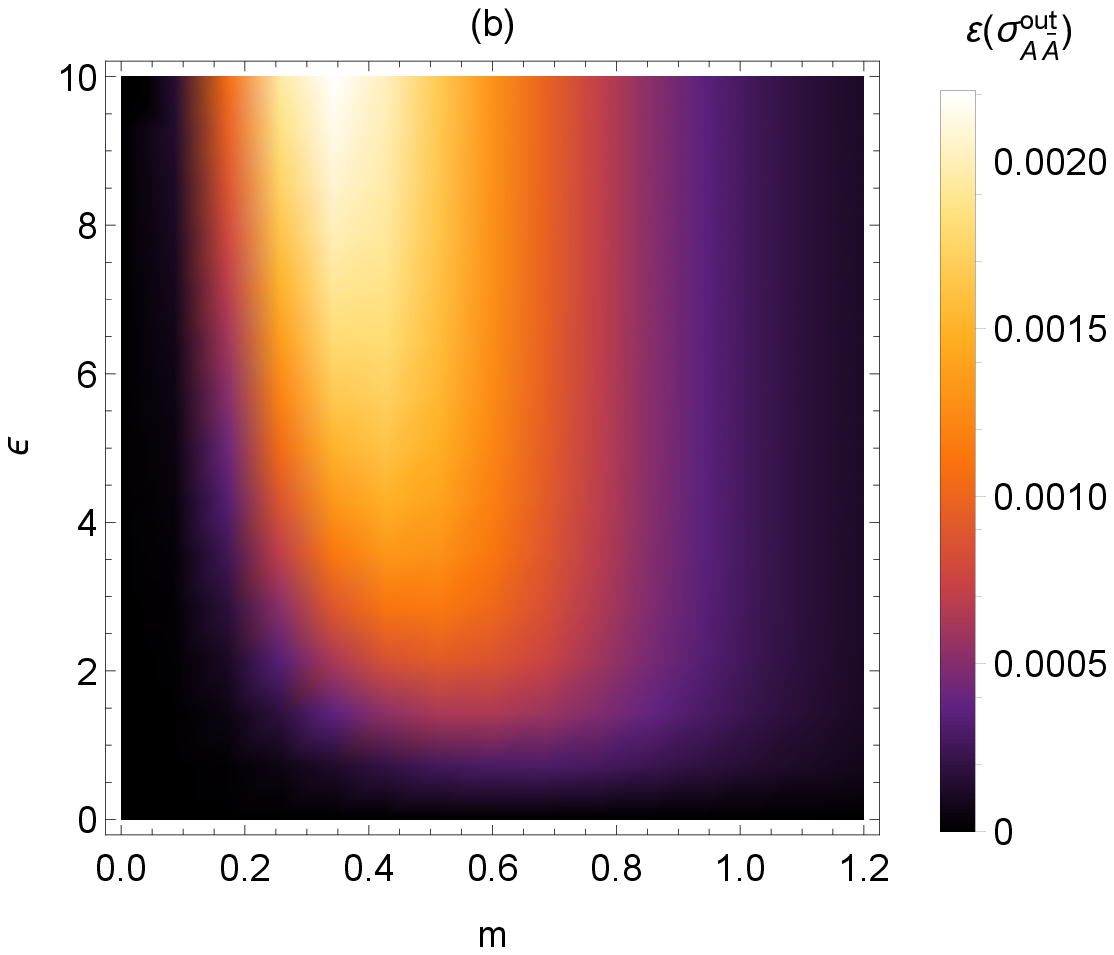}
\label{fig1c}
\end{minipage}%
\caption{(a) Quantum entanglement $\mathcal{E}(\sigma_{A\bar A}^{\rm out})$ between the mode $A$ and the mode $\bar A$ as functions of the expansion rate $\sigma$ and the momentum $k$ for the fixed values of $m=\epsilon=s=1$. (b) Quantum entanglement $\mathcal{E}(\sigma_{A\bar A}^{\rm out})$ as functions of the mass $m$ and the expansion volume $\epsilon$ for the fixed values of $k=\sigma=s=1$.}
\label{Fig2}
\end{figure}

In Fig.\ref{Fig2} (a), we plot the Gaussian entanglement $\mathcal{E}(\sigma_{A\bar A}^{\rm out})$ as functions of the expansion rate $\sigma$ and the momentum $k$.
Fig.\ref{Fig2} (b) shows how the mass $m$ and the expansion volume $\epsilon$
influence the Gaussian entanglement $\mathcal{E}(\sigma_{A\bar A}^{\rm out})$.
Fig.\ref{Fig2} (a) shows that the Gaussian entanglement $\mathcal{E}(\sigma_{A\bar A}^{\rm out})$ increases with the expansion rate $\sigma$, which means that  the expansion of the universe can generate quantum entanglement between Alice and anti-Alice. We find that the
particles with the smaller momentum $k$ help the expansion of the universe to
generate greater quantum entanglement. From Fig.\ref{Fig2} (b), we can see that quantum entanglement $\mathcal{E}(\sigma_{A\bar A}^{\rm out})$  increases with the growth of the expansion volume $\epsilon$. When we choose the particles with the suitable mass $m$, the expansion of the underlying spacetime has a more pronounced effect on quantum entanglement. From Fig.\ref{Fig2}, we can also see that quantum entanglement is sensitive to the expansion rate but not to the expansion volume.

Combining Fig.\ref{Fig1} and Fig.\ref{Fig2}, we come to three conclusions: (i) with the increase of the expansion rate and the expansion volume, the initial Gaussian entanglement between Alice and Bob decreases, and at the same time, the Gaussian entanglement between Alice and anti-Alice (or Bob and anti-Bob) increases, which means that the expansion of the universe redistributes the initial entanglement; (ii) quantum entanglement is more sensitive to the expansion rate than the expansion volume; (iii) the information about the expanding universe  can be better extracted  by choosing the particles with the smaller momentum and the appropriate mass.
Studying the properties of quantum entanglement can help us to
understand the history of the expanding universe. Therefore, these results can help us better simulate the  expanding spacetime and the production of cosmological particles with quantum systems in a laboratory setting   \cite{oL45m,oL45m1,oL45m1po}.

\subsection{Generation of a four-mode Gaussian entanglement}
The four-mode Gaussian state $\sigma_{AB \bar A \bar B}^{\rm out}$
of Eq.(\ref{w21}) is  entirely inseparable, meaning that it contains genuine multipartite Gaussian entanglement distributed among all the four parties involved. Let us now compute the bipartite entanglements in the $1\rightarrow3$ partitions of the Gaussian state $\sigma_{AB \bar A \bar B}^{\rm out}$. Employing Eq.(\ref{w5}), the $1\rightarrow3$ entanglements are found to be
\begin{equation}\label{w28}
    \mathcal{E}(\sigma_{A|B\bar A\bar B}^{\rm out})=\ln \bigg\{\frac{\cosh(2s)+\theta_k^2}{1-\theta_k^2}  \bigg\},
\end{equation}
\begin{equation}\label{w29}
    \mathcal{E}(\sigma_{\bar A|AB\bar B}^{\rm out})=\ln \bigg\{\frac{\cosh(2s)+\theta_k^2}{1-\theta_k^2}  \bigg\}.
\end{equation}
For any nonzero value of $s$, $\sigma$ and $\epsilon$, each single party is in an entangled Gaussian state with the block of the remaining three parties, with respect to all possible global splitting of the modes.

\begin{figure}
\begin{minipage}[t]{0.5\linewidth}
\centering
\includegraphics[width=3.0in,height=5.2cm]{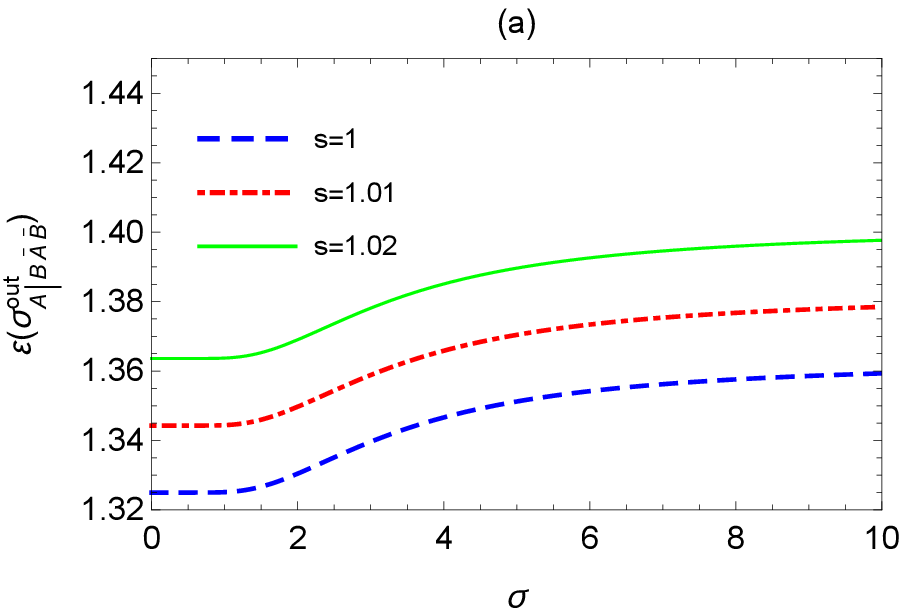}
\label{fig1a}
\end{minipage}%
\begin{minipage}[t]{0.5\linewidth}
\centering
\includegraphics[width=3.0in,height=5.2cm]{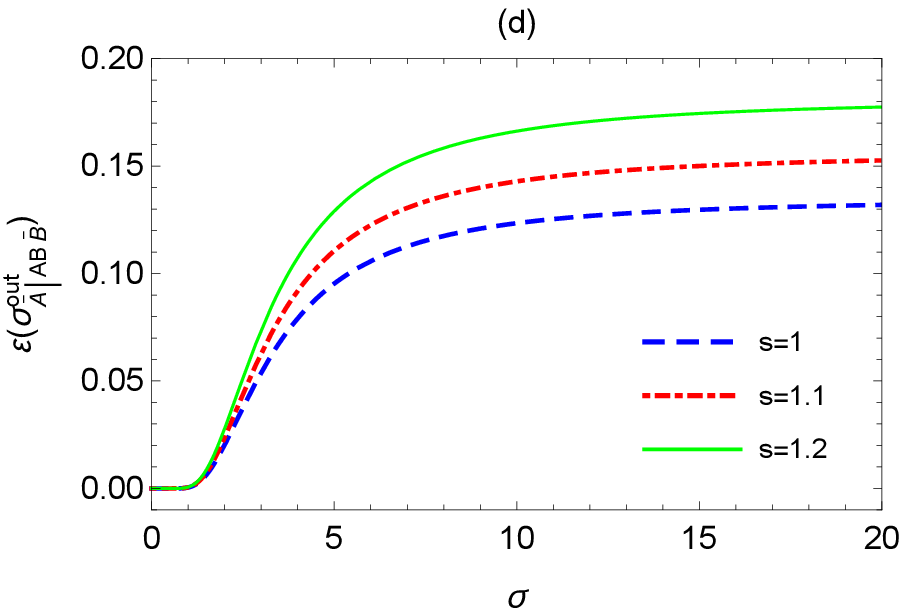}
\label{fig1a}
\end{minipage}%

\begin{minipage}[t]{0.5\linewidth}
\centering
\includegraphics[width=3.0in,height=5.2cm]{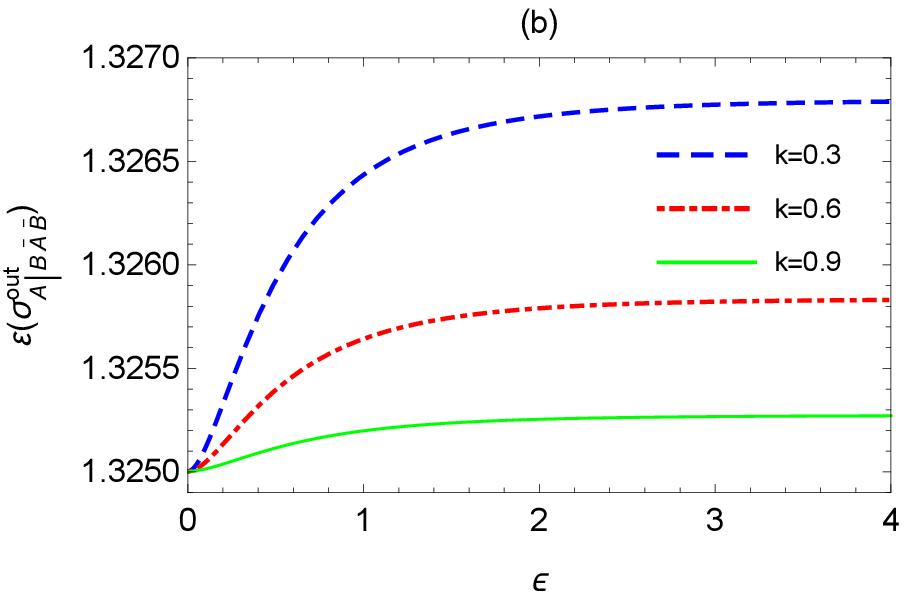}
\label{fig1c}
\end{minipage}%
\begin{minipage}[t]{0.5\linewidth}
\centering
\includegraphics[width=3.0in,height=5.2cm]{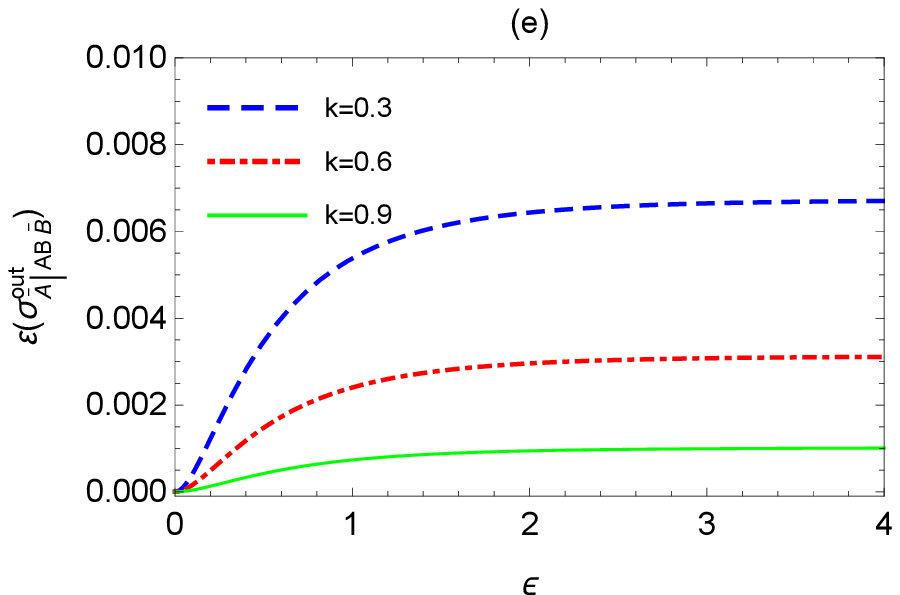}
\label{fig1a}
\end{minipage}%

\begin{minipage}[t]{0.5\linewidth}
\centering
\includegraphics[width=3.0in,height=5.2cm]{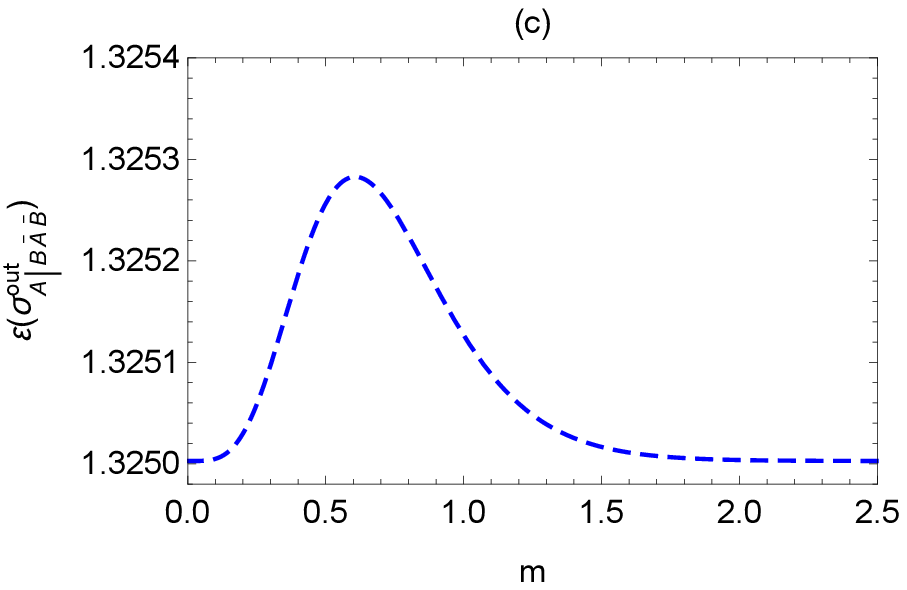}
\label{fig1c}
\end{minipage}%
\begin{minipage}[t]{0.5\linewidth}
\centering
\includegraphics[width=3.0in,height=5.2cm]{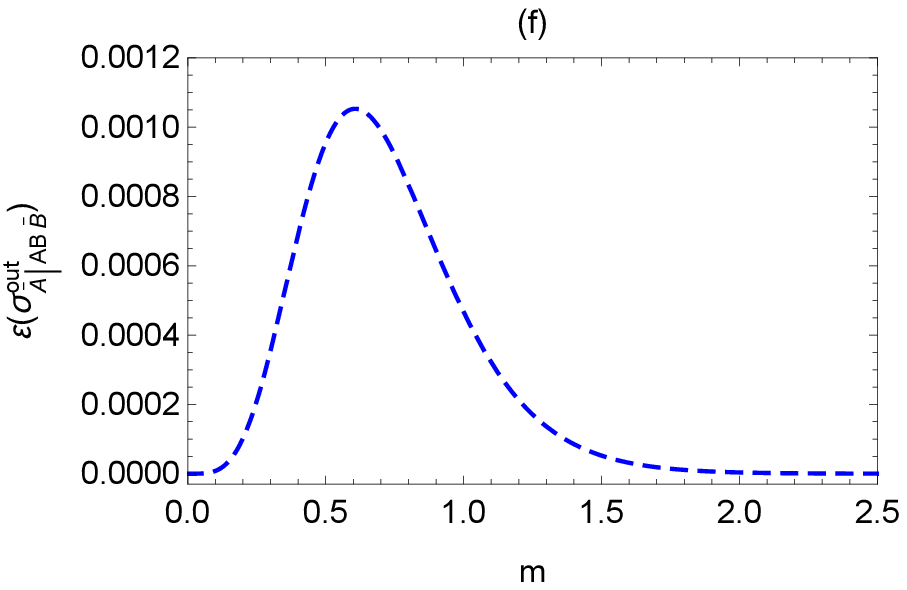}
\label{fig1a}
\end{minipage}%
\caption{$1\rightarrow3$ entanglement (a) and (d)  as a function of the expansion rate $\sigma$ for different squeezing parameters $s$. $1\rightarrow3$ entanglement (b) and (e)  as a function of the expansion volume $\epsilon$ for different momentums $k$. $1\rightarrow3$ entanglement (c) and (f)  as a function of the mass $m$. The other parameters are set to $1$.}
\label{Fig3}
\end{figure}

In Fig.\ref{Fig3} (a) and (d), we plot the $1\rightarrow3$ bipartite entanglement as  a function of the expansion rate $\sigma$ for different squeezing parameters $s$.
We find that the $1\rightarrow3$ bipartite entanglement monotonically increases
with the increase of the expansion rate $\sigma$.
In addition, as the squeezing parameter $s$ increases, the influence of the expansion rate $\sigma$ on the generated entanglement $\mathcal{E}(\sigma_{\bar A|AB\bar B}^{\rm out})$ is more obvious.

The $1\rightarrow3$ bipartite entanglement is plotted in Fig.\ref{Fig3} (b) and (e)
as a function of the expansion volume $\epsilon$  for different momentums $k$.
It is shown  that the $1\rightarrow3$ bipartite entanglement increases
with the increase of the expansion volume $\epsilon$. However, the influence of the expansion volume on the $1\rightarrow3$ entanglement is less than that of the expansion rate.
From Fig.\ref{Fig3} (b) and (e), we can see that, for a fixed expansion volume $\epsilon$,
 the $1\rightarrow3$ entanglement decreases with the increase of the  momentum $k$.

 Fig.\ref{Fig3} (c) and (f) show how the mass $m$
influences the $1\rightarrow3$ entanglement. From Fig.\ref{Fig3} (c), we find that
quantum entanglement $\mathcal{E}(\sigma_{A|B\bar A\bar B}^{\rm out})$ first increases from the initial entanglement to the maximum and then reduces to the initial entanglement. From
Fig.\ref{Fig3} (f), we can see that quantum entanglement $\mathcal{E}(\sigma_{\bar A|AB\bar B}^{\rm out})$ increases from zero to the maximum and then reduces to zero.
This implies that, for the bosons with the larger mass $m$, the effect of the expansion of the universe on  the $1\rightarrow3$ entanglement is not apparent.

The residual multipartite entanglement  from
the monogamy inequality in Eq.(\ref{w6}) is an entanglement monotone
under Gaussian local operations and classical communication for a four-mode  pure state $\sigma_{AB \bar A \bar B}^{\rm out}$.
The residual  entanglement of the Gaussian state $\sigma_{AB \bar A \bar B}^{\rm out}$
is defined as
\begin{eqnarray}\label{w30}
   && \mathcal{E}^{res}(\sigma_{AB\bar A\bar B}^{\rm out})=\rm min\bigg\{\mathcal{E}(\sigma_{A|B\bar A\bar B}^{\rm out})-\mathcal{E}(\sigma_{AB}^{\rm out})-\mathcal{E}(\sigma_{A\bar A}^{\rm out})-\mathcal{E}(\sigma_{A\bar B}^{\rm out}),\\ \nonumber &&\mathcal{E}(\sigma_{\bar A|AB\bar B}^{\rm out})-\mathcal{E}(\sigma_{\bar AA }^{\rm out})-\mathcal{E}(\sigma_{\bar AB}^{\rm out})-\mathcal{E}(\sigma_{\bar A\bar B}^{\rm out})\bigg\}.
\end{eqnarray}
We  verify that the second quantity can achieve the minimum. Therefore, we obtain an analytic expression for the residual entanglement $\mathcal{E}^{res}(\sigma_{AB\bar A\bar B}^{\rm out})$
\begin{eqnarray}\label{w31}
   \mathcal{E}^{res}(\sigma_{AB\bar A\bar B}^{\rm out})=\ln \bigg\{\frac{\theta_k^2\cosh(2s)+1}{1+\theta_k^2}  \bigg\}.
\end{eqnarray}
From Eq.(\ref{w31}), we find that the residual entanglement $\mathcal{E}^{res}(\sigma_{AB\bar A\bar B}^{\rm out})$ is always greater than or equal to zero for any  parameters, which proves that the Coffman-Kundu-Wootters inequality in Eq.(\ref{w6}) is still correct in an expanding spacetime. In addition, it precisely quantifies  the multipartite quantum correlations that cannot be stored in the bipartite form.

\begin{figure}
\begin{minipage}[t]{0.5\linewidth}
\centering
\includegraphics[width=3.0in,height=5.2cm]{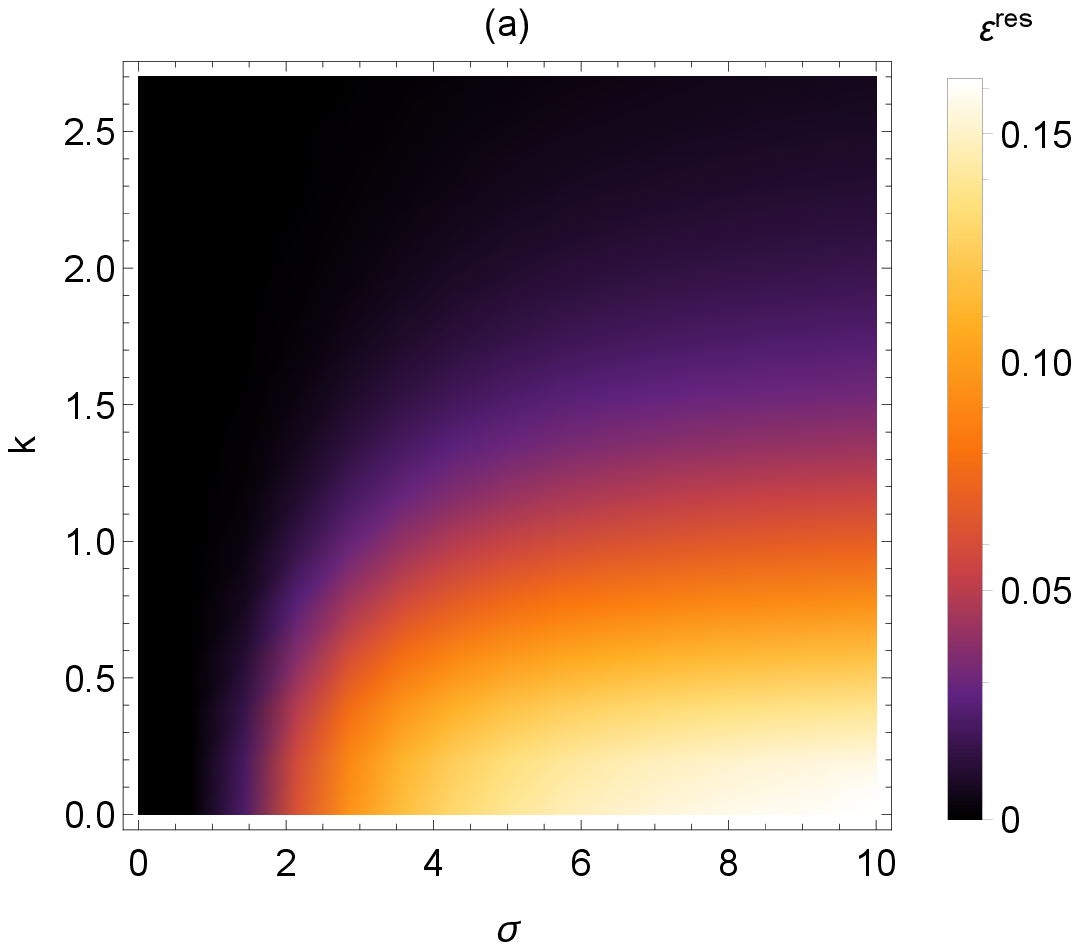}
\label{fig1a}
\end{minipage}%
\begin{minipage}[t]{0.5\linewidth}
\centering
\includegraphics[width=3.0in,height=5.2cm]{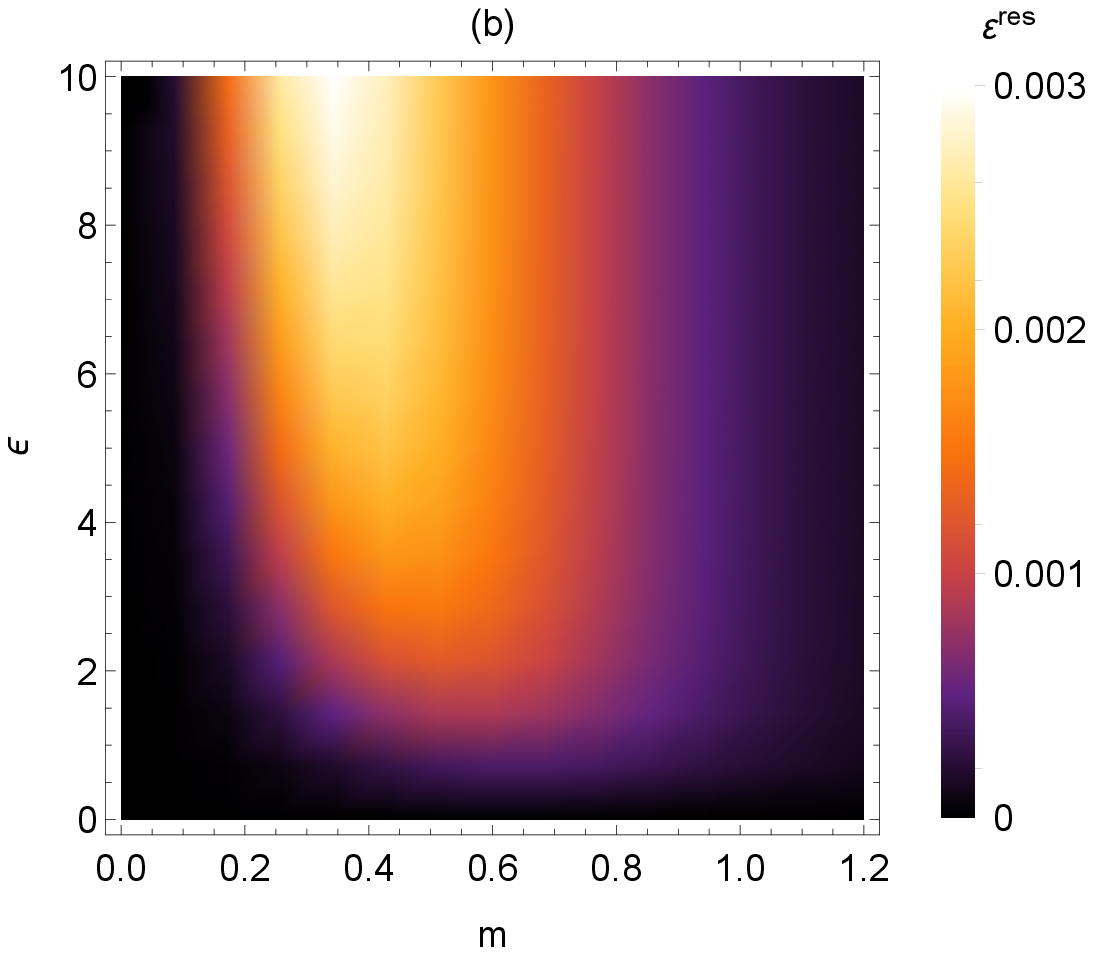}
\label{fig1c}
\end{minipage}%
\caption{(a) The residual entanglement $\mathcal{E}^{res}(\sigma_{AB\bar A\bar B}^{\rm out})$ as functions of the expansion rate $\sigma$ and the momentum $k$ for the fixed values of $m=\epsilon=s=1$. (b) The residual entanglement $\mathcal{E}^{res}(\sigma_{AB\bar A\bar B}^{\rm out})$ as functions of the mass $m$ and the expansion volume $\epsilon$ for the fixed values of $k=\sigma=s=1$.}
\label{Fig4}
\end{figure}

In Fig.\ref{Fig4} (a), we plot the residual entanglement $\mathcal{E}^{res}(\sigma_{AB\bar A\bar B}^{\rm out})$  as  functions of the expansion rate $\sigma$ and the momentum $k$.
Fig.\ref{Fig4} (b) shows how the mass $m$ and the expansion volume $\epsilon$
influence the residual entanglement $\mathcal{E}^{res}(\sigma_{AB\bar A\bar B}^{\rm out})$.
From Fig.\ref{Fig4}, we can see that the residual entanglement monotonically increases with the increase of the expansion rate and the expansion volume. This indicates that the expansion of the universe can generate the residual entanglement. We can also see that
the residual entanglement is more influenced by the expansion rate than the expansion volume. In addition, the
residual entanglement is more sensitive to the cosmological parameters when we choose the particles with  the smaller momentum and the suitable mass  in an expanding spacetime.

\section{ Conclutions  \label{GSCDGE}}
We have studied the redistribution of quantum entanglement for continuous variables
in an expanding spacetime.
We consider four modes: the bosonic mode $A$ observed by Alice; the bosonic mode $B$ observed by Bob; the antibosonic  mode $\bar A$ observed by anti-Alice; the antibosonic mode  $\bar B$ observed by anti-Bob.
We get a phase space description for a quantum state evolution under the influence of the expansion of a Robertson-Walker spacetime. When the quantum state evolves from the asymptotic past to the asymptotic future, the dynamical  entanglement contains  historical information concerning the expanding spacetime.
We find that quantum entanglement is more sensitive to the expansion rate than the expansion volume. We show the redistribution of the initial entanglement: quantum entanglement between the modes $A$ and $B$
decreases with the growth of the expansion rate and the expansion volume; at the same time,
quantum entanglement between the modes $A$ and $\bar A$ (or $B$ and  $\bar B$ ) can be generated by the expansion of the underlying spacetime.
This means that the loss of quantum entanglement can be interpreted as a redistribution of the initial  entanglement into multipartite quantum correlations.
We also find that the $1\rightarrow 3$ entanglement increases with the expansion rate and the expansion volume. The residual entanglement is always greater than or equal to zero, which directly proves that the Coffman-Kundu-Wootters inequality  is true in an expanding spacetime. According to the analysis of quantum entanglement,
choosing the particles with the smaller momentum and the optimal mass is a better way to extract information about the expanding universe.
These results can guide the simulation of the expanding universe  in different quantum systems \cite{oL45m,oL45m1,oL45m1po}.

\begin{acknowledgments}
This work is supported by the National Natural
Science Foundation of China (Grant Nos. 12205133, 1217050862, 11275064, 11975064 and 12075050 ), LJKQZ20222315 and 2021BSL013.	
\end{acknowledgments}

$\textbf{Data Availability Statement}$

This manuscript has no associated data.

$\textbf{Conflict of interest}$

The authors declare no conflicts of interest.



\begin{thebibliography}{99}

\bibitem{L1}
C. H. Bennett, G. Brassard, C. Cr\'{e}peau, R. Jozsa, A. Peres,
and W. K. Wootters, Phys. Rev. Lett. {\bf70}, 1895 (1993).

\bibitem{L2}
S. F. Huegla, M. B. Plenio, and J. A. Vaccaro, Phys. Rev. A
{\bf65}, 042316 (2002).

\bibitem{L3}
J. L. Dodd, M. A. Nielsen, M. J. Bremner, and R. T. Thew,
Phys. Rev. A {\bf65}, 040301 (2002).

\bibitem{L4}
The Physics of Quantum Information, edited by D. Bouwmeester,
A. Ekert, A. Zeilinger (Springer-Verlag, Berlin,
2000).


\bibitem{L5}
I. Fuentes-Schuller, R. B. Mann, Phys. Rev. Lett. {\bf95}, 120404 (2005).

\bibitem{L6}
P. M. Alsing, I. Fuentes-Schuller, R. B. Mann and T. E. Tessier,
Phys. Rev. A {\bf74}, 032326 (2006).


\bibitem{LA1}
D. E. Bruschi, J. Louko, E. Mart\'{\i}n-Mart\'{\i}nez, A. Dragan, and I. Fuentes,
Phys. Rev. A {\bf82}, 042332 (2010).

\bibitem{LA2}
D. E. Bruschi, A. Dragan, I. Fuentes, and J. Louko, Phys. Rev. D {\bf86}, 025026 (2012).

\bibitem{L7}
Y. Li, Q. Mao and Y. Shi, Phys. Rev. A {\bf99}, 032340 (2019).

\bibitem{L8}
Q. Pan and J. Jing,
Phys. Rev. D {\bf78}, 065015 (2008).

\bibitem{L9}
M. R. Hwang, D. K. Park, and E. Jung,
Phys. Rev. A {\bf83}, 012111 (2011).

\bibitem{L10}
J. Wang, and J. Jing,
 Phys. Rev. A {\bf83}, 022314 (2011).

\bibitem{L11}
S. M. Wu, Y. T. Cai, W. J. Peng, H. S. Zeng, Eur. Phys. J. C {\bf82}, 412 (2022).


\bibitem{L12}
Y. Dai, Z. Shen, and Y. Shi,
Phys. Rev. D {\bf94}, 025012 (2016).

\bibitem{L13}
W. C. Qiang, G. H. Sun, Q. Dong, and S. H. Dong,
 Phys. Rev. A {\bf98}, 022320 (2018).

\bibitem{L14}
S. M. Wu, H. S. Zeng, Eur. Phys. J. C {\bf82}, 4 (2022).

\bibitem{L15}
X. Liu, Z. Tian, J. Wang, J. Jing, Phys. Rev. D {\bf97}, 105030 (2018).

\bibitem{L16}
Z. Tian, S. Y. Ch\"{a}, U. R. Fischer, Phys. Rev. A {\bf97}, 063611 (2018).

\bibitem{L17}
J. Wang, C. Wen, S. Chen, J. Jing,  Phys. Lett. B {\bf800}, 135109 (2020).

\bibitem{L18}
S. Kanno, J. P. Shock, and J. Soda, Phys. Rev. D {\bf94}, 125014 (2016).

\bibitem{L19}
A. Albrecht, S. Kanno and M. Sasaki, Phys. Rev. D {\bf97}, 083520 (2018).

\bibitem{L20}
S. M. Wu, H. S. Zeng, T. Liu, New J. Phys. {\bf24}, 073004  (2022).


\bibitem{OL1}
A. J. Torres-Arenas, Q. Dong, G. H. Sun, W. C. Qiang, S. H. Dong,
Phys. Lett. B {\bf789}, 93 (2019).

\bibitem{OL2}
S. Xu, X. K. Song, J. D. Shi, and L. Ye,
Phys. Rev. D {\bf89}, 065022 (2014).

\bibitem{OL3}
A. Matsumura and Y. Nambu, Phys. Rev. D {\bf98}, 025004 (2018).

\bibitem{OL4}
S. M. Wu, H. S. Zeng, Eur. Phys. J. C {\bf82}, 716 (2022).

\bibitem{OL5}
Z. Tian, L. Wu, L. Zhang, J. Jing, J. Du, Phys. Rev. D {\bf106}, L061701  (2022).

\bibitem{OL6}
D. Wang, F. Ming, X. K. Song, L. Ye,  J. L. Chen, Eur. Phys. J. C {\bf80}, 800 (2020).

\bibitem{OL7}
S. Bhattacharya, H. Gaur, N. Joshi, Phys. Rev. D {\bf102}, 045017 (2020).

\bibitem{OL8}
S. M. Wu, H. S. Zeng, H. M. Cao, Class. Quantum Grav. {\bf38}, 185007 (2021).

\bibitem{OL9}
L. J. Li, F. Ming, X. K. Song, L. Ye, D. Wang, Eur. Phys. J. C {\bf82}, 726 (2022).

\bibitem{OL10}
S. Bhattacharya, N. Joshi, Phys. Rev. D {\bf105}, 065007 (2022).

\bibitem{LDF1}
S. M. Wu, C. X. Wang, D. D. Liu, X. L. Huang, H. S. Zeng,  J. High Energy Phys. {\bf02}, 115 (2023).

\bibitem{LDF2}
D. E. Bruschi, I. Fuentes, and J. Louko, Phys. Rev. D {\bf85}, 061701(R) (2012).

\bibitem{LDF3}
N. Friis, D. E. Bruschi, J. Louko,  I. Fuentes, Phys. Rev. D {\bf85}, 081701(R) (2012).

\bibitem{LDF4}
D. E. Bruschi, J. Louko, D. Faccio, I. Fuentes, New J. Phys. {\bf15}, 073052 (2013).

\bibitem{LDF5}
A. Dragan, J. Doukas, E. Mart\'{\i}n-Mart\'{\i}nez,  D. E. Bruschi, Class. Quantum Grav. {\bf30}, 235006  (2013).




\bibitem{L21}
D. Ahn, Y. Moon, R. Mann, and I. Fuentes-Schuller, J.
High Energy Phys. {\bf06}, 062 (2008).

\bibitem{L22}
L. Bombelli, R. K. Koul, J. Lee, and R. D. Sorkin, Phys.
Rev. D {\bf34}, 373 (1986).

\bibitem{L23}
G. T. Horowitz and J. Maldacena, J. High Energy Phys. {\bf02},
008 (2004) .

\bibitem{L24}
S. Lloyd, Phys. Rev. Lett. {\bf96}, 061302 (2006).

\bibitem{L25}
J. L. Ball, I. Fuentes-Schuller, F. P. Schuller, Phys. Lett. A
{\bf359}, 550 (2006).

\bibitem{L26}
E. Mart\'{\i}n-Mart\'{\i}nez, N. C. Menicucci, Class. Quantum Grav.
{\bf29}, 224003 (2012).

\bibitem{L27}
X. Liu, J. Jing, J. Wang, Z. Tian, Quantum Inf Process {\bf19}, 26 (2020).

\bibitem{L28}
I. Fuentes, R. B. Mann, E. Mart\'{\i}n-Mart\'{\i}nez and S. Moradi,
Phys. Rev. D {\bf82}, 045030 (2010).

\bibitem{L29}
A. Einstein, B. Podolsky, N. Rosen, Phys Rev {\bf47}, 777 (1935).

\bibitem{L30}
S. L. Braunstein, P. van Loock,  Rev Mod Phys {\bf77}, 513 (2005).

\bibitem{oL45m1}
J. Steinhauer, $et$ $al$, Nature Communications {\bf13}, 2890 (2022).


\bibitem{oL45m}
Z. Tian, J. Jing, A. Dragan, Phys. Rev. D {\bf95}, 125003 (2017).


\bibitem{oL45m1po}
C. Viermann, $et$ $al$, Nature {\bf611}, 260 (2022).



\bibitem{L31}
D. Buono, G. Nocerino, A. Porzio, and S. Solimeno, Phys.
Rev. A {\bf86}, 042308 (2012).

\bibitem{L32}
Q. Y. He, Q. H. Gong, and M. D. Reid, Phys. Rev. Lett. {\bf114}, 060402 (2015).

\bibitem{L33}
G. Adesso, A. Serafini, F. Illuminati, Phys. Rev. A {\bf70}, 022318  (2004).

\bibitem{L34}
R\'{e}nyi A 1960 On measures of information and entropy Proc. 4th Berkeley Symposium onMathematics, Statistics
and Probability pp 547-61.

\bibitem{L35}
M. Headrick, Phys. Rev. D {\bf82}, 126010 (2010).

\bibitem{L36}
G. Adesso, D. Girolami, A. Serafini, Phys. Rev. Lett. {\bf109}, 190502 (2012).

\bibitem{L37}
G. Adesso, S. Ragy and A. R. Lee, Open Syst. Inf. Dyn. {\bf21}, 1440001 (2014).

\bibitem{L38}
C. M. Caves and B. L. Schumaker, Phys. Rev. A {\bf31}, 3068
(1985).

\bibitem{L39}
B. L. Schumaker and C. M. Caves, Phys. Rev. A {\bf31}, 3093 (1985).

\bibitem{L40}
G. Adesso, I. Fuentes-Schuller, and M. Ericsson, Phys. Rev.
A {\bf76}, 062112 (2007).

\bibitem{L41}
G. Adesso, S. Ragy and D. Girolami, Class. Quantum Grav.  {\bf29}, 224002 (2012).








\end{thebibliography}
\end{document}